\documentclass[fleqn,10pt]{wlscirep}
\usepackage[utf8]{inputenc}
\usepackage[T1]{fontenc}
\usepackage{array} 
\newcommand{\overbar}[1]{\mkern 1.5mu\overline{\mkern-1.5mu#1\mkern-1.5mu}\mkern 1.5mu}
\title{The evolution of knowledge within and across fields in modern physics}

\author[1]{Ye Sun}
\author[1,2,*]{Vito Latora}
\affil[1]{School of Mathematical Sciences, Queen Mary University of London, London E1 4NS, United Kingdom}
\affil[2]{Dipartimento di Fisica e Astronomia, Università di Catania and Istituto Nazionale di Fisica, I-95123 Catania, Italy}

\affil[*]{v.latora@qmul.ac.uk}

\begin{abstract}
The exchange of knowledge across different areas and disciplines plays a key
role in the process of knowledge creation, and can stimulate
innovation and the emergence of new fields. We develop here a
quantitative framework to extract significant dependencies among
scientific disciplines and turn them into a 
time-varying network whose nodes 
are the different fields, while the weighted links
represent the flow of knowledge from one field to another at a given
period of time.
Drawing on a comprehensive data set on scientific production in modern
physics and on the patterns of citations between articles published in
the various fields in the last thirty years, we are then able to map, over time,
how the ideas developed in a given field in a certain time period 
have influenced later discoveries in the same field or in
other fields. 
The analysis of knowledge flows internal to each field displays a
remarkable variety of temporal behaviours, with some fields of physics
showing to be more self-referential than others. 
The temporal networks of knowledge exchanges across fields
reveal cases of one field continuously absorbing knowledge 
from another field in the entire observed period, 
pairs of fields mutually influencing each other, but also cases of
evolution from absorbing to mutual or even to back-nurture
behaviors.
\end{abstract}
\begin{document}

\flushbottom
\maketitle
%
%
\thispagestyle{empty}

\section*{Introduction}
Knowledge creation and knowledge sharing go hand in
hand.
Knowledge is in fact created through combination and integration of
different concepts, and can benefits from social interactions and
interdisciplinary collaborations.
Recent works have explored from many angles how 
knowledge flows across scholars~\cite{sekara2018chaperone, li2019early,monechi2019efficient, armano2017beneficial,milojevic2018changing}, institutions~\cite{clauset2015systematic,gargiulo2014driving,deville2014career,ma2015anatomy} and disciplines~\cite{van2015interdisciplinary,sinatra2015century,battiston2019taking}.
In particular, it has been shown that knowledge exchange across fields can influence
the evolution of culture and language~\cite{bhagat2002cultural,chen2010impact}, strengthen multi-faceted cooperation~\cite{bell2007geography,monechi2019efficient}, and drive the innovation and development of science~\cite{sorenson2006complexity,agrawal2008spatial,meyer2002tracing,acemoglu2016innovation}.
Research publications are one of the primary channels of communication
for the exchange and spreading of knowledge in science~\cite{zeng2017science}. By publishing
their own articles and citing works by their peers, researchers
continuously contribute to the processes of knowledge creation,
knowledge sharing and knowledge acquisition~\cite{zhang2013characterizing}, thereby promoting the advancement of science. The presence of a citation between two 
research articles often denotes a certain transfer of knowledge from
the cited article to the citing articles. 
It is therefore natural to use citations  
between articles published in different scientific fields to 
investigate the flow of knowledge across different
domains of science.  
Despite some works in this direction have already started elucidating 
the main mechanisms of knowledge sharing
and diffusion~\cite{borner2006mapping,zhuge2002knowledge,yan2016disciplinary},
a systematic study on how knowledge evolves in time 
and of the complex interactions and influences
between different fields is still lacking.

In this article, we propose a novel framework to detect and quantify
relevant transfers of knowledge across disciplines and between 
different time periods. One of the outcomes of the method is 
the construction of a time-varying network mapping the structure
of knowledge and the relations between disciplines.
In particular, we present an application to study the evolution of 
scientific knowledge in modern physics, namely to investigate how influences
from one field of physics to another have evolved over
time in the last thirty years. Building on bibliographic information of over $430,000$ articles
published by the American Physical Society (APS) between $1985$ and
$2015$, and making use of the highest-level Physics and Astronomy
Classification Scheme (PACS) codes, which indicate the fields of
physics an article belong to, we construct a temporal network where
the nodes represent the fields of modern physics and the directed
links denote the presence of a significant dependence of a field from
another. Such a network is changing over time and, as we will show below,
its analysis by the methods of network science~\cite{latora2017complex, newman2018networks} is able to reveal
essential properties of how knowledge is exchanged among fields and
over different time periods. We have found that, overall,   
knowledge flows have become increasingly
homogeneous over the last years, indicating the important role of
interdisciplinary
research~\cite{pan2012evolution,van2015interdisciplinary,sinatra2015century,bonaventura2017advantages,pluchino2019exploring}.
In spite of this, some typical patterns of influence, such as cases of 
one field absorbing knowledge from another field, or two  
fields mutually influencing each other, 
clearly emerge at the microscopic scale. 
Our findings provide insights into the basic mechanisms of knowledge 
exchange in science, and can turn very useful to understand the dynamics of
scientific production and the growth of novelties in scientific
domains~\cite{tria2014dynamics,iacopini2018innovation,chinazzi2019mapping}. 

\section*{Results}
\textbf{The fields of modern physics.} An overall insight into the main fields of modern physics can be
obtained by a basic analysis of the characteristics of the APS data
sets.  Considered at their highest level, PACS codes divide modern
physics into ten major fields (see Table ~\ref{PACS}). A measure of
the relevance of each field can then be derived from the volume of
papers published in each field. Since each paper can be listed with
multiple PACS codes we assign it to multiple fields.  We therefore
consider each paper as one unit of knowledge and define the field
composition of the paper as the relative frequency of its PACS
codes. For instance, if a paper is listed with three PACS codes
$\textsl{89.75.-k}$, $\textsl{81.05.-t}$ and $\textsl{05.45.-a}$, we
assign two-thirds of this paper to \textsl{Interdisciplinary
Physics} (PACS $80$), and the remaining one-third to \textsl{General
Physics} (PACS $00$).  The total number of papers $N_{paper}$
associated to each field over the entire time period of $31$ years is
reported in Table~\ref{PACS}. One can see that the three largest
fields are \textsl{Condensed Matter} (PACS $60$ and $70$) and
\textsl{General Physics} (PACS $00$), capturing $57\%$ of the entire
publications in APS journals. \textsl{GPE} is the smallest field, with
only $8325$ papers, which is roughly one-fifteenth the size of the
largest field \textsl{CM$2$}. To quantify and compare the growth rate
of each field, the average yearly change 
in the number of papers $\Delta N_{paper}$ 
is also reported in Table~\ref{PACS}. 
Consistently with the rankings based on 
field sizes, \textsl{GEN} and \textsl{CM$2$} also exhibit the
highest growth rate, larger than $100$ papers per year, while 
\textit{GPE} shows the slowest increase, with only $5$ papers per year on
average. On the contrary, \textsl{CM$2$}, the third-largest field by size,
is ranked fourth from the bottom according to the average
growth $\Delta N_{paper}$. 
An opposite trend is observed for
\textsl{Interdisciplinary Physics} and \textsl{Astrophysics}, which 
respectively take fifth and sixth place according to their growth 
growth rates, although their field sizes are ranked eighth and ninth
among these fields, reflecting their rapid development during the observing period.

We found that $91\%$ of the papers have more than one PACS code, with
$36\%$ of them reporting PACS codes which are at least from two
different fields. To quantify the level of interdisciplinarity of a
given field, we have collected all the papers with at least one PACS code
from that field, and then calculated 
the proportion $J$ of these papers which are also classified by at least
one PACS code from other fields. The results in Table~\ref{PACS}
show that \textsl{Interdisciplinary Physics} is the field with the largest
value of $J$: almost $90\%$ of the 
Interdisciplinary Physics papers are also classified by PACS
codes from other fields of physics. This result is 
consistent with the expectation that interdisciplinary research
combines knowledge from various disciplines. 
Instead, papers in the fields of \textsl{Nuclear},
\textsl{Particles} and \textsl{Condensed matter $2$} physics 
are more likely to use PACS codes from their own fields. Summing up, 
the above analyses indicate that the differences between
fields of physics are remarkable, either in terms of the size and growth
of the fields, and in terms of their interactions with other fields.

\begin{table*}[h!]
\renewcommand\arraystretch{1.2}
\centering
\begin{tabular}{p{1cm}<{\centering} p{1.5cm}<{\centering} p{8cm}p{1cm}<{\centering} p{1cm}<{\centering}p{1cm}<{\centering}}
 \hline
 \hline
  PACS  & Abbreviation & Field Information &$N_{paper}$& $\Delta N_{paper}$ &$J$ \\
 \hline
00 & GEN   &General Physics        & 66909  & 115 & 0.76  \\
10 & EPF  &The Physics of Elementary Particles and Fields  & 46722  & 56  & 0.44 \\
20 & NUC   & Nuclear Physics    & 29120 & 17 &0.42  \\
30 & ATM  &Atomic and Molecular Physics & 28929  & 10 &0.64  \\
40 & EOA  &Electromagnetism, Optics, Acoustics, Heat Transfer, Classical Mechanics, Fluid Dynamics & 35425 & 58 & 0.79\\
50 & GPE  &Physics of Gases, Plasmas, Electric Discharges  & 8325 & 5 &0.62 \\
60 & CM1 &Condensed Matter: Structural, Mechanical and Thermal Properties & 53287 & 21 &0.78 \\
70 & CM2   &Condensed Matter: Electronic Structure, Electrical, Magnetic, Optical Properties  & 127319 & 106& 0.46\\
80 & IPR  &Interdisciplinary Physics and Related Areas of Science and Technology& 24346 & 51 &0.89 \\
90 & GAA  &Geophysics, Astronomy, Astrophysics & 15319  & 34 & 0.78  \\
 \hline
 \hline
\end{tabular}
\caption {The ten fields of modern physics. {\rm PACS codes and names of the main fields of physics as defined at the highest level of the APS hierarchical classification scheme. $N_{paper}$ represents the total number of papers published in each field in the period between 1985 and 2015. $\Delta N_{paper}$ denotes the average yearly increase in the number of papers in each field. 
Among all the papers in an observed field, $J$ indicates the proportion of these papers that are also classified with at least one PACS code from other fields.}}
 \label{PACS}
\end{table*}

\bigbreak
\noindent
\textbf{The knowledge flow network.} Interactions among scientific fields can be better characterized
by making use of scientific citations.   
A published article in a scientific field citing articles of
another field implies the cited field reflects a piece of previously
existing knowledge that the citing field builds upon. And this, in
turn, indicates a flow of knowledge from the cited field to the
citing field. Hence, we can construct a network of knowledge flow
across fields by analyzing the pattern of citations among 
papers of different fields. 
The nodes of such a network
represent the ten fields of modern physics as indicated by the
PACS codes, while the directed links between fields 
denote the flows of knowledge from one area of physics to another. 
\begin{figure*}
\centering
    \includegraphics[width=14cm]{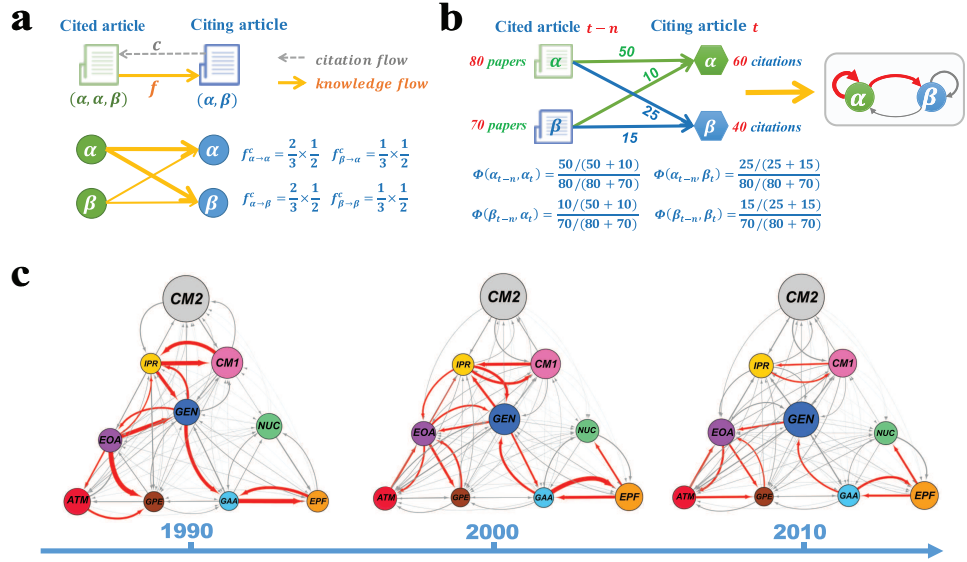}
    \caption{\textbf{The knowledge flow network and its time evolution.}
      (a) Illustration of how a citation 
      between two papers is translated into a contribution to the knowledge flow 
      between the two corresponding fields.
      (b) Construction of a weighted network of
      knowledge flow based on the significance of each
      link. (c) The knowledge flow network among different fields
      in physics in years $1990$,
      $2000$ and $2010$. Node sizes are proportional to the number of
      papers published in each field and given year, and the line
      widths correspond to the weights of knowledge flows between two
      fields. The links with weights larger than $1$ are selected with
      red color. The arrow represents the direction of knowledge
      flows.}
    \label{fig:MethodNetwork}
\end{figure*}
Specifically, for a given citation ${c}$ there will be a transfer of
knowledge from each PACS code in the cited reference to all the PACS
codes in the citing article.  We hence indicate as
$f^{c}_{\alpha  \rightarrow \beta}$ 
the volume of knowledge flow from field $\alpha$
to field $\beta$ due to citation $c$. As shown in
Fig.~\ref{fig:MethodNetwork}(a), this is calculated as the
product of the proportion of PACS codes from field $\beta$ in the
citing article times the proportion of PACS codes from field $\alpha$
in the cited article. This ensures the
normalization $\sum_{\alpha}\sum_{\beta}{f}^{c}_{\alpha
  \rightarrow \beta}=1$ for each citation $c$,
meaning that each citation contributes a unit of knowledge
transfer that is then split to the different fields.  
For instance, in Fig.~\ref{fig:MethodNetwork}(a), two of the three 
PACS codes of the cited article belong to field $\alpha$, while 
one over two of the PACS codes in the citing article is from field $\beta$.
Consequently, we assume that the volume of knowledge
flowing from field $\alpha$ to field $\beta$, due to citation $c$, is 
$f^{c}_{\alpha \rightarrow \beta}=2/3 \times 1/2$.
Similarly, we can calculate the quantities
$f^{c}_{\alpha \rightarrow \alpha}$, $f^{c}_{\beta \rightarrow \alpha}$ and $f^c_{\beta
  \rightarrow \beta}$. 
In order to characterize the flow of knowledge across fields and to
study its evolution over the years, we construct yearly
aggregated networks by selecting different pairs of years for  
citing and cited articles respectively. This is done by 
analyzing all the citations from papers
published in a given year $t$ to papers published in year $t-n$, 
and defining the total volume of knowledge flowing from field $\alpha$ to
field $\beta$ as: 
\begin{equation}
  F_{\alpha \rightarrow \beta}^{t-n \rightarrow t} =
  \sum_{\rm{cit} } f^{c}_{\alpha \rightarrow \beta}
\label{WeightCitations}
\end{equation}
where the sum runs over all citations $c$ from papers published in
field $\beta$ in year $t$ to papers published in 
field $\alpha$ in year $t-n$.
Notice that $n$ is a tunable parameter, denoting the relative age of cited
papers with respect to the citing year $t$.
Having the possibility to vary both $t$ and $n$ allows to take into
account that the probability of a citation
is influenced by: {\em (i)} the relative 
age of the two papers~\cite{zhu2003effect}, and by
{\em (ii)} the number of papers published in
the cited year $t-n$.
%
The quantities $F_{\alpha \rightarrow \beta}^{t-n \rightarrow t}$ are, 
however, affected by field-specific
characteristics and publishing conventions, such as typical field sizes 
and time-varying growth rates, which, as shown in Table~\ref{PACS},
may vary a lot from field to field.
Hence, an increase of $F_{\alpha \rightarrow \beta}^{t-n \rightarrow t}$ over
time does not automatically reflect a closer relation between fields
$\alpha$ and $\beta$, as it can only be due to a rapid growth in the number
of publications in these two fields. 
In order to account for this, we define the statistical
significance $\phi(\alpha_{t-n},\beta_{t})$, which 
quantifies how 
the observed knowledge flow $F_{\alpha \rightarrow \beta}^{t-n \rightarrow t}$ 
exceeds the flow expected in a opportunely chosen null model (See the Method section).
Fig.~\ref{fig:MethodNetwork}(b) illustrates an example 
of how to calculate the quantities $\phi(\alpha_{t-n},\beta_{t})$
in Eq.(\ref{Independent}). 
Suppose, for instance, field $\alpha$ in year $t-n$ 
provides a total of $75$ units of knowledge to all the fields
in year $t$, with field $\alpha$ itself receiving $50$ of these
$100$ units, and $\beta$ getting $25$,
i.e. $F_{\alpha \rightarrow \alpha}^{t-n \rightarrow t}=50$ and
$F_{\alpha \rightarrow \beta}^{t-n \rightarrow t}=25$.
Analogously we assume field $\beta$ in year $t-n$ provides
a total of $25$ units to year $t$, $15$ to field $\beta$ itself
and $10$ to $\alpha$.  
Now, the marginal probability that the citing field 
is $\beta$ can be obtained as 
$Pr(X_{t}^{citing}=\beta)=\sum_{\alpha} F_{\alpha \rightarrow
  \beta}^{t-n \rightarrow t}/\sum_{\alpha,\beta} F_{\alpha
  \rightarrow \beta}^{t-n \rightarrow t}=40/(60+40)$, while 
$Pr(Y_{t-n}^{cited}=\alpha| X_{t}^{citing}=\beta)= F_{\alpha
  \rightarrow \beta}^{t-n \rightarrow t}/\sum_{\alpha} F_{\alpha
  \rightarrow \beta}^{t-n \rightarrow t} = 25/(25+15)$.
We therefore have $P(\alpha_{t-n}, \beta_{t}) = 25/40 \cdot 40/100$.
Such a probability needs to be compared to that of a null
model in which $P^{\rm rand}(\alpha_{t-n}, \beta_{t})=Pr(Y_{t-n}^{cited}=\alpha) \cdot Pr(X_{t}^{citing}=\beta)= 80/150 \cdot 40/100$, since the probability $Pr(Y_{t-n}^{cited}=\alpha)$ that the cited paper in year $t-n$ is in field $\alpha$ is equal to $80/(80+70)$. Finally, the ratio $\phi(\alpha_{t-n},\beta_{t})$
in Eq.~(\ref{Independent}) is equal to $25/40 \cdot 150/80$.

Analogously, we can calculate the statistical significance of all the other flows reported in Fig.~\ref{fig:MethodNetwork}(b).
Such quantities allow to capture the intrinsic
variation of knowledge flows among fields and also to compare different
pair of fields. 
Finally, to obtain the weights of knowledge flows in year $t$ from a
cited time window $\Delta t'$, we define the flow weights $w_{\alpha
  \rightarrow \beta}^{\Delta t' \rightarrow t}$ for each couple of
cited field $\alpha$ and citing field $\beta$ as:
\begin{equation}
	w_{\alpha \rightarrow \beta}^{\Delta t' \rightarrow t}= \frac{1}{\vert\Delta t' \vert}\sum_{n \in \Delta t'}{\phi (\alpha_{t-n}, \beta_{t})}
\label{eq:sigwindow1}
\end{equation}
where $\vert\Delta t' \vert$ is the length of the time window. Let
$\Delta t'=[1, 5]$ and $\vert \Delta t' \vert=5$, then one can
construct the significant knowledge flow network in each year $t$ from
previous $5$ years. Furthermore, for each given source period $\Delta
t'$, one can also investigate the knowledge flows within an observing
period $\Delta t$:
\begin{equation}
	w_{\alpha \rightarrow \beta}^{\Delta t' \rightarrow \Delta t}= \frac{1}{\vert \Delta t \vert}\sum_{t \in \Delta t} w_{\alpha \rightarrow \beta}^{\Delta t' \rightarrow t}
\label{SigWindow2}
\end{equation}
For example, let $\vert \Delta t \vert=5$, we can divide the entire
time period into five observing time windows, namely $[1990, 1994]$,
$[1995, 1999]$, $[2000, 2004]$, $[2005, 2009]$ and $[2010, 2014]$. The
weight of each link in the network reflects how significant the
knowledge flows between two related fields. This quantitative
framework enable us to investigate the evolution of knowledge flows in
two time dimensions: (i) for each given observing period $\Delta t$,
the weights of knowledge flows from different time interval $\Delta
t'$ can be observed; and also (ii) for each fixed $\Delta t'$, the
weights of knowledge flows within different observing period $\Delta
t$ can be compared. 

\bigbreak
\noindent
\textbf{Temporal analysis of knowledge flow networks.} We first investigate how the overall properties of the knowledge
flow networks have changed over time. Specifically we have evaluated,
for each year, the flows of knowledge from the previous $5$ years, i.e
we have fixed $\Delta t'=[1,5]$ and $\vert \Delta t \vert=1$ in our
framework. To better visualize the temporal changes, the whole
knowledge flow networks obtained for the three years $1990$, $2000$
and $2010$ are reported in Fig.~\ref{fig:MethodNetwork}(c).
Links representing a significant flow of knowledge ($w_{\alpha
\rightarrow \beta}^{\Delta t' \rightarrow t}>1$) are shown in red
color. 
\begin{figure}[htb]
\centering
    \includegraphics[width=14cm]{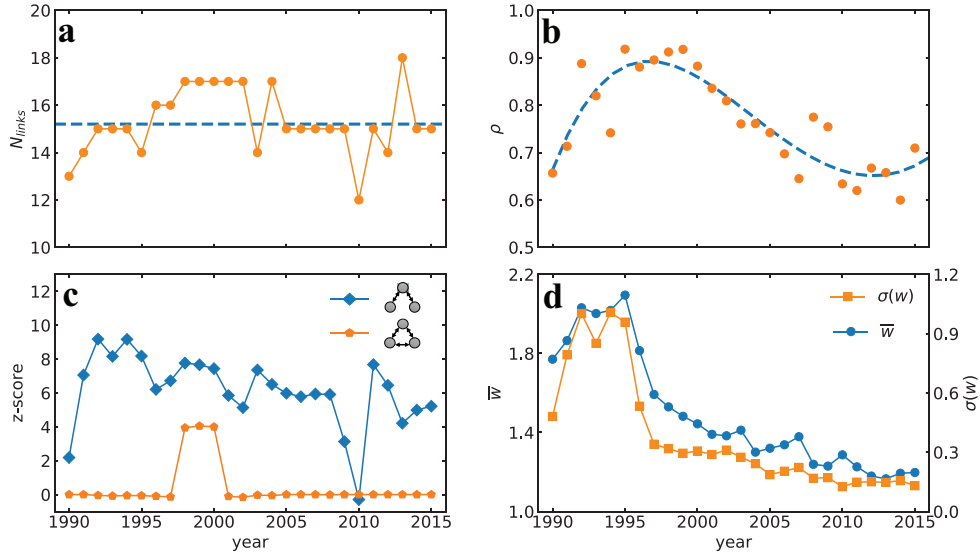}
    \caption{\textbf{Temporal analysis of knowledge flow networks.}
      (a) The number of significant links in the knowledge flow network
      is shown as a function of the year  
      together with its time average, reported as a dashed gray line.
      (b) Network reciprocity measuring the proportion of
      bidirectional links, shows a pattern
      with a peak around year $1998$.
      The top $50\%$ of bidirectional links with the largest
      sum of mutual weights were considered in the computation of
      the reciprocity. (c) The Z-score of two types of three-node motifs
      is reported as a function of time.
      (d) Mean and standard deviation of the weights of
      significant links gradually decrease over time, indicating that
      the knowledge flows between fields is tending more towards 
      random expectations.}
    \label{fig:TemporalAnalysis}
\end{figure}
The first thing to notice is that the number of significant
links is roughly constant over the years, as
also illustrated in Fig.~\ref{fig:TemporalAnalysis}(a).
In addition to this, we observe that more
links are reciprocated in $2000$ with respect to years 
$1990$ and $2010$, which suggests that situation in which couples
of fields mutually influence each other are more common 
$2000$. To further examine this, we have computed the network
reciprocity (see Materials and Methods) for each year.  
The results reported in Fig.~\ref{fig:TemporalAnalysis}(b)
indicate that the value of the reciprocity $\rho$ has increased 
in the first few years, reached a peak around $1998$,
and then has begun to decrease in the
following years. This has lead us to conclude that the  
highest levels of mutuality in knowledge transfer among different fields
of physics have been experienced between $1995$ and $2000$.

We have then extracted the typical patterns of knowledge transfer
in the network. For this reason, we have focused on the statistically
significant three-node motifs in the knowledge flow networks
\cite{milo2002network}, i.e. the directed
connected subgraphs of three nodes that appear in the network
more often than they would occur by chance. 
Fig.~\ref{fig:TemporalAnalysis}(c) illustrates the
Z-scores (See the Method section) of two relevant three-node motifs
over the years. One can see that the subgraph represented by bi-directed
paths (diamond symbol) is the most significant motif throughout the whole
time period, with a Z-score on average equal to about $6$. 
Furthermore, complete subgraphs of three nodes, corresponding to three mutually
connected fields of physics, are only statistically significant in the
period from $1998$ to $2000$, where the complete subgraph  
\textsl{GEN}, \textsl{EOA} and \textsl{IPR} appears. Notice 
that this period also corresponds to the time period of
high reciprocity in Fig.~\ref{fig:TemporalAnalysis}(b).

In addition, Fig.~\ref{fig:MethodNetwork}(c) indicates 
that there are fewer links with large weights in year $2010$ than in
$1990$ and $2000$. To further investigate this trend,
Fig.~\ref{fig:TemporalAnalysis}(d) reports 
mean $\overbar{w}$ and standard deviation
$\sigma_w$ of the weights of 
significant links between $1990$ and
$2016$. We find that both the 
values of $\overbar{w}$ and $\sigma_{w}$ gradually decrease
overtime and eventually stabilize to values slightly above $1$ and $0$ respectively.
This indicates that the exchange of knowledge across domains has 
increasingly become more homogeneous with respect to 
the beginning of 1980s, when each field only absorbed 
knowledge from a handful of close domains. From a different 
perspective, this also reflects a rise of the interdisciplinary
character of research in physics.

\begin{figure}
\centering
    \includegraphics[width=14cm]{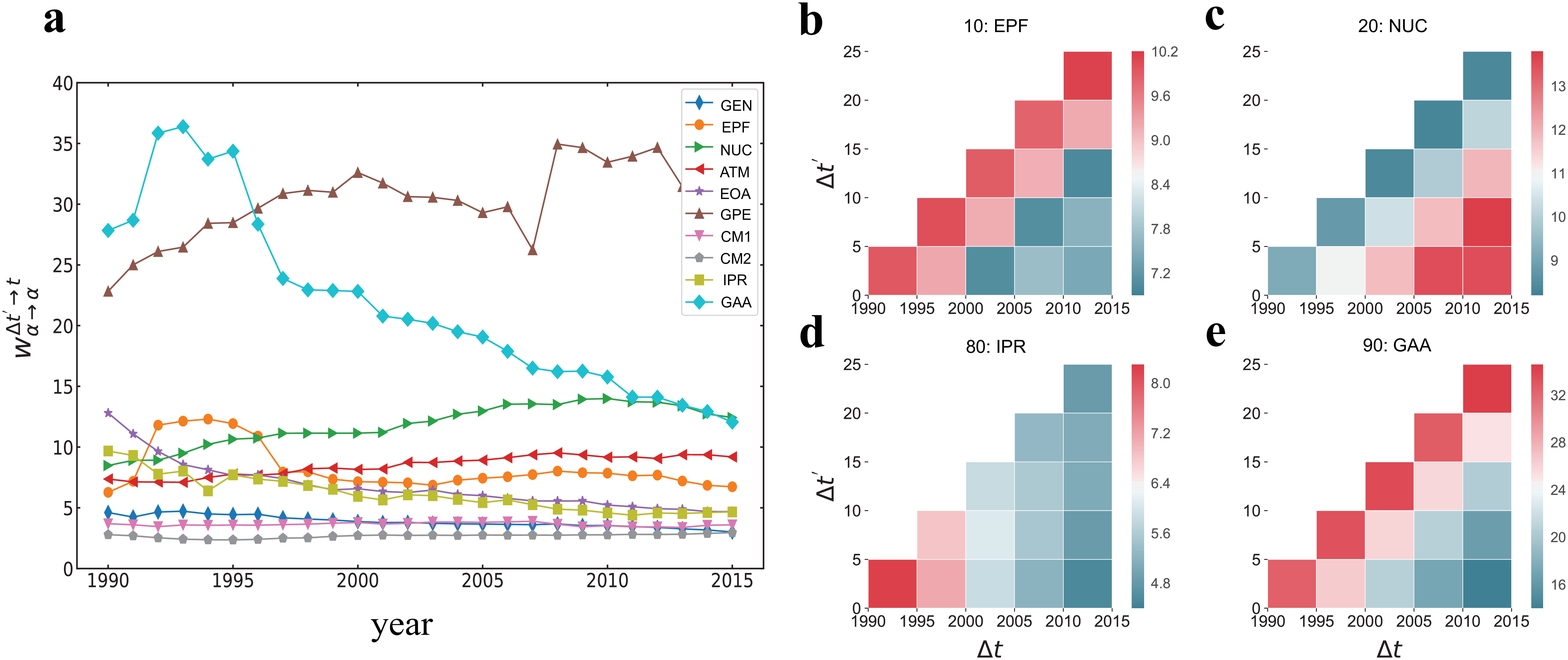}
    \caption{\textbf{Evolution of knowledge flows within a field}  (a) For
      each field $\alpha$ and each year $t$, we plot the internal
      knowledge flow $w_{\alpha \rightarrow \alpha}^{\Delta t' \rightarrow t}$  
      from a window of the previous $5$ years to $t$.  
      In (b)-(e) we show the evolution of internal
      flows for four specific fields in two-dimensional plots $\Delta
      t$, $\Delta t'$. By comparing the change in each row, we find
      that field \textsl{NUC} shows an increasingly high degree of
      self-reference over time, while, conversely, \textsl{IPR} and
      \textsl{GAA} tend to become less and less
      self-dependent. Focusing on the variation in each column, we can
      examine the effect of reference age on the significance of
      internal knowledge transfer. The lengths of each citing period
      $\Delta t$ and cited period $\Delta t'$ in (b)-(e) are both
      equal to $5$ years.}
    \label{fig:Internal}
\end{figure}

\bigbreak
\noindent
\textbf{Internal knowledge flows.} The weights of the internal flows $w_{\alpha \rightarrow \alpha}^{\Delta t' \rightarrow t}$ from a field $\alpha$ to itself
are an indication of the degree of self-dependence of the research
field. To investigate the evolution of the internal
knowledge flows, we have computed, for each of the ten fields of physics,
the weights of the internal flows in every observing year $t$ (between $1990$ and $2015$) from each
of the previous five years, namely adopting a cited time window $\Delta t'=[1,5]$.
Fig.~\ref{fig:Internal}(a) indicates that the internal
knowledge flows are significant ($w_{\alpha \rightarrow
  \alpha}^{\Delta t' \rightarrow t}>1$) for all ten fields over the
whole $21$-year period of time, although the temporal trends can vary
from field to field. The two fields with the largest variations are
\textsl{GAA} and \textsl{GPE}.  Field \textsl{GAA} exhibits a
remarkable decrease in the degree of self-reference after $1993$,
indicating that in this field the internal transfer of knowledge has
become less and less significant over time. Conversely, field \textsl{GPE} shows
an increasing trend and becomes the most self-referential field after
$1995$. Other fields exhibit decreasing (\textsl{EOA} and \textsl{IPR}) 
or increasing (\textsl{NUC} and \textsl{ATM}) patterns, 
while the contribution of internal flows are 
relatively low and keeps nearly constant for fields such as \textsl{GEN},
\textsl{CM1} and \textsl{CM2}.

In Fig.~\ref{fig:Internal}(b-e) we focus on the evolution
of internal flows of the four fields \textsl{EPF}, \textsl{NUC}, \textsl{IPR}
and \textsl{GAA}. In particular, we perform a two-dimensional analysis in which
we change the positions of both the observing time windows $\Delta t$ and the source
time window $\Delta t'$. We consider the case where the lengths of the two
time windows is the same and is equal to $5$ years. The colours
in Fig.~\ref{fig:Internal}(b-e) represent
the values of internal knowledge flows
$w_{\alpha \rightarrow \alpha}^{\Delta t' \rightarrow \Delta t}$. 
By looking at the variations of colours in each row we find that field 
\textsl{NUC} shows an increasingly high degree of self-reference over
time, while \textsl{IPR} and \textsl{GAA} tend to lower their  
degree of internal flows, which is consistent with the results in
Fig.~\ref{fig:Internal}(a).

By looking at the variation
of colours over each column of Fig.~\ref{fig:Internal}(b-e) 
we can instead investigate the influence of reference's age
on the internal flows of knowledge. One can see that fields such as \textsl{NUC} 
and \textsl{IPR} show a decreasing trend 
from most recent times to the past, in agreement with previous studies stating that the likelihood of
a paper being discovered significantly decreases with the papers' age
~\cite{tahamtan2016factors}. By contrast, we observe an unexpected and very clear pattern  
for \textsl{EPF} and \textsl{GAA}, since both fields exhibit a maximum of the values
along the anti-diagonal line. Notice that each square along the anti-diagonal
line represents the same cited time window, namely a time window of five
years before the period $[1990,1994]$. 

This may be due to important discoveries and the publication of pioneering
research works in the fields \textsl{EPF} and \textsl{GAA} during the period $[1985,1990]$  
which would clearly increase the probability for researchers in the field 
to cite, in the following years, papers published in that period.
A possible explanation can be for instance the rapid development 
in the period $[1985,1989]$ of the new research area "\textsl{astroparticle physics}",   
emerging at the intersection of particle physics, astronomy and
astrophysics~\cite{cirkel2008history}, 
and which mainly combines the knowledge from fields \textsl{EPF} and \textsl{GAA}.
As an evidence of this rapid development, notice
that even a new journal named "Astroparticle Physics" was established in $1992$.
Moreover, the fact that the weights of the internal flows in
\textsl{GAA} are nearly three times larger than those in \textsl{EPF},
can be due to the Hubble Space Telescope, one of the major scientific
breakthroughs in field \textsl{GAA}. The telescope is one of the
largest and most productive scientific research tool for astronomy,
and it was indeed launched in $1990$ (within the period of interest in
the anti-diagonal line), greatly promoting the development of astronomy
in \textsl{GAA}.

\begin{figure*}
\centering
    \includegraphics[width=16cm]{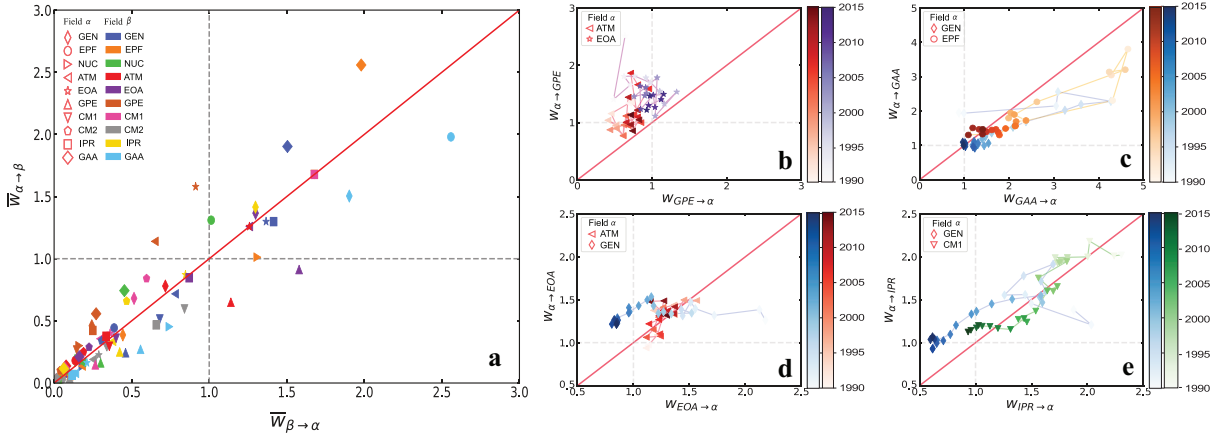}
    \caption{\textbf{Evolution of knowledge across fields}.  (a) For
      each pair of fields we plot the average flows of knowledge in
      either direction, averaged over the entire observation period of
      26 years. (b)-(e) report some of the typical patterns of
      temporal evolution we have observed over the years.  Symbol
      color (from light to dark) indicates the years from $1990$ to
      $2015$, while the lines join consecutive years to help following
      the trajectories.  The bisector red line corresponds to the case
      of perfectly symmetric knowledge flows between the two
      fields. (b) "Absorbing mode": field \textit{GPE} has been
      absorbing knowledge from fields \textit{ATM} and \textit{EOA}
      throughout the entire time period. (c) "From absorbing to
      mutual mode": \textit{GEN} and \textit{EPF} have initially
      absorbed more knowledge from field \textit{GAA} and then tend to
      a balanced case in which they absorb from \textit{GAA} the same
      knowledge they provide to it. (d) "Back-nurture mode": while
      during the first few years \textit{GEN} has absorbed more
      knowledge from field \textit{EOA} than it has contributed to, at
      a later stage the situation is inverted.  (e) "Mutual mode":
      field \textit{IPR} and \textit{CM1} tend to share knowledge in
      a symmetric way over the whole period.
      }
    \label{fig:symmetry}
\end{figure*}

\bigbreak
\noindent
\textbf{The evolution of knowledge flows across fields.} Examining how the discoveries in a field have contributed to a
different field of physics is even more important than studying the
flows of knowledge within a given field. In order to get an overall
picture of the existing influences across different fields of modern
physics, we report in Fig.~\ref{fig:symmetry}(a) the average
weights of knowledge flows between each couple of fields over the
whole period under study. To highlight the mutual exchange of flows,
the results are shown in a ($\overbar{w}_{\alpha \rightarrow
  \beta}$)-($\overbar{w}_{\beta \rightarrow \alpha}$) plane. Each
point refers to a pair of fields, and the distance from the position
of the point to the bisector (red line) measures the level of asymmetry
in the exchange of knowledge between the two fields. 
We notice that most of the points are concentrated around the bisector,
especially those points in the lower-left corner corresponding to 
pairs of fields with small significance weights. However, there are
also points far from the line, such as the point corresponding to the pair 
\textit{GPE} and \textit{ATM} (red
up-triangle in the lower right of the panel), indicating asymmetric
transfers of knowledge between two fields.
To investigate the temporal evolution in the exchange of 
knowledge between two fields in Fig.~\ref{fig:symmetry}(b-e), we have considered the same types
of plots over time. In such a case, each pair of fields
corresponds to a trajectory joining the points corresponding to the
different years from $1990$ to $2015$. 
The colors of symbols
from light to dark indicates the years from past to the most
recent. Although the significance of the links
in general decreases over time, the temporal patterns can
vary from one pair of fields to another. The four panels illustrate 
the four major classes of behaviour (modes) we have found, namely:
absorbing, absorbing to mutual, back-nurture and mutual mode. 
Absorbing mode can be seen in field \textit{GPE}, which has
absorbed more knowledge from fields \textit{ATM} and \textit{EOA}
throughout the whole period under study
(Fig.~\ref{fig:symmetry}(b)). Fields \textit{GEN} and \textit{EPF} show a similar behavior as $GPE$ in the beginning, absorbing more knowledge
from \textit{GAA}, while in the last few years, \textit{GEN} and
\textit{EPF} tend to mutually exchange knowledge with  \textit{GAA}, although the weights on the links in both directions become less
significant 
(Fig.~\ref{fig:symmetry}(c)). More interestingly, we also find
a back-nurture mode as shown in
Fig.~\ref{fig:symmetry}(d). Field \textit{GEN} at first absorbs
more knowledge from \textit{EOA} than what it provides to
\textit{EOA}, but later the situation is inverted. 
Finally, fields \textit{IPR} and \textit{CM1} shows another pattern, the mutual mode, indicating that they have exchanged knowledge in an almost symmetric way over the whole period. These different evolution patterns clearly demonstrate that the processes of knowledge creation and transfer across fields can be highly heterogeneous.  

\section*{Discussion}

Knowledge sharing and transfer across scientific disciplines, and 
cross-fertilization are increasingly recognized as crucial factors to
breakthrough innovation in science~\cite{rinia2002measuring,phene2006breakthrough,battiston2019taking}.
The temporal network approach we have proposed in this article can be useful to shed lights on the evolution of knowledge within a field and on the dynamic patterns of influences between different fields. Our study case application 
has shown that major developments in physics
can influence a field for many decades and can even trigger knowledge production in other fields. Indeed, the patterns of cross-fertilization 
vary greatly among the different disciplines of physics and can also
show marked transitions over time. For instance, the physics of
gases and plasmas has consistently absorbed knowledge from 
atomic and molecular physics and from electromagnetism 
over the last three decades. Other fields such as condensed matter
and interdisciplinary physics have instead always 
shared and mutually exchanged knowledge. Finally, we have
revealed interesting transitions from absorbing to mutual modes,
for instance in the case of the physics of elementary particles,
a field of physics that has initially been strongly influenced by
astronomy and astrophysics, but in the new century has
also contributed to the progress of these latter disciplines.
Our findings not only shed new lights on the basic laws governing the
development of scientific fields, but can also have practical implications on
the future development of economic
policies and research strategies.

\section*{Methods}
\small
\textbf{Data.} The data set contains $435,717$ articles published by the American Physical Society (APS) from year $1985$ until the end of $2015$. 
Publication date, Physics and Astronomy Classification
Scheme (PACS) codes, and bibliography have been extracted for each article. 
The PACS codes are grouped
into a five-level hierarchy and each of them indicates a very specific
field of physics. As an example, the PACS code $\textsl{64.60.aq}$,
indicating the field \emph{"Networks"}, belongs to the
broader field \emph{"Equations of state, phase equilibria, and phase
  transitions"} (PACS $64$) and further belongs to top-level field
\emph{"Condensed Matter: Structural, Mechanical and Thermal
  Properties"}(PACS $60$). Here, we consider the PACS codes at the
highest level, which classify the physics into ten main fields (Table
~\ref{PACS}). Each article is associated with up to four PACS
codes. With regard to the bibliography, only the citations referring
to articles published in the APS journals were considered.

\bigbreak
\noindent
\textbf{Null model and statistically significant networks.} To characterize the flow of knowledge across fields and at different
time periods, the statistical significance of each contribution has
been validated with respect to an appropriately chosen null model.
For each couple of fields all the citations from papers
published in citing year $t$ to papers published in cited year
$t-n$ have been considered. 
Let ${\rm X}_{t}^{\rm citing}$ be the field of citing papers
published in year $t$, and ${\rm Y}_{t-n}^{\rm cited}$ the field of
cited papers published in year $t-n$.  We indicate as $P
(\alpha_{t-n}, \beta_{t}) = {\rm Pr}({\rm Y}_{t-n}^{\rm cited}=\alpha,
{\rm X}_{t}^{\rm citing}=\beta)$ the joint probability that papers
published in year $t$ in field $\beta$ cite papers published in year
$t-n$ in field $\alpha$.  Such a probability can be written as:
\begin{equation}
  P(\alpha_{t-n}, \beta_{t})= {\rm Pr}({\rm Y}_{t-n}^{\rm cited}=\alpha| {\rm X}_{t}^{\rm citing}=\beta) \times
  {\rm Pr}({\rm X}_{t}^{\rm citing}=\beta)
\label{MassFunction}
\end{equation}
where $ {\rm Pr}({\rm Y}_{t-n}^{\rm cited}=\alpha| {\rm X}_{t}^{\rm
  citing}=\beta)$ is the conditional probability of ${\rm
  Y}_{t-n}^{\rm cited}=\alpha$ given that ${\rm X}_{t}^{\rm
  citing}=\beta$, and $ {\rm Pr}({\rm X}_{t}^{\rm citing}=\beta) $ is
the marginal probability.
We then consider a null model in which the papers published in year $t$
in field $\rm X$ randomly select papers published in year $t-n$ as their
citations, regardless of which fields they belong to. Hence, the joint probabilities in the null model can be written in terms of the marginal probabilities as: 
\begin{equation}
	P^{\rm rand}(\alpha_{t-n}, \beta_{t})={\rm Pr}({\rm Y}_{t-n}^{\rm cited}=\alpha) \cdot {\rm Pr}({\rm X}_{t}^{\rm citing}=\beta)
\label{Independent}
\end{equation}
By calculating the ratio $\phi (\alpha_{t-n}, \beta_{t})$ between the two 
probabilities in Eq.~(\ref{MassFunction}) and (\ref{Independent}):
\begin{equation}
	\phi (\alpha_{t-n}, \beta_{t})= \frac{ {\rm Pr}({\rm Y}_{t-n}^{\rm cited}=\alpha|{\rm X}_{t}^{\rm citing}=\beta)}{{\rm Pr}({\rm Y}_{t-n}^{\rm cited}=\alpha)}
\label{Significance}
\end{equation}
we were able to quantify how the observed flows of knowledge deviate from the flows expected to arise simply from random choices.
A value $\phi(\alpha_{t-n},\beta_{t})=1$ has been adopted as critical threshold
to distinguish whether the knowledge flow from field $\alpha$ 
to field $\beta$ is statistically 
significant, with $\phi(\alpha_{t-n},\beta_{t})>1$ indicating that
field $\beta$ in year $t$ is more likely to have extracted knowledge
from field $\alpha$ in year $t-n$ than would be expected at random.

\bigbreak
\noindent
\textbf{Network analysis.} To characterize the networks of knowledge flow we have evaluated the network
reciprocity and we have performed a motif analysis. 
\bigbreak
\noindent
\emph{(i) Reciprocity.} For each year $t$ we have computed the reciprocity coefficient~\cite{garlaschelli2004patterns}
of the knowledge flow network as:
\begin{equation}
	\rho_t=\frac{\sum\limits_{\alpha \neq \beta} (w_{\alpha \rightarrow \beta}^{\Delta t' \rightarrow t}-\overbar{w}_{\Delta t' \rightarrow t})(w_{\beta \rightarrow \alpha}^{\Delta t' \rightarrow t}-\overbar{w}_{\Delta t' \rightarrow t})}{\sum\limits_{\alpha \neq \beta}(w_{\alpha \rightarrow \beta}^{\Delta t' \rightarrow t}-\overbar{w}_{\Delta t' \rightarrow t})^2}
\label{eq:Reciprocity}
\end{equation}
where the average value $\overbar{w}_{\Delta t' \rightarrow t} \equiv
\sum_{\alpha \neq \beta}w_{\alpha \rightarrow \beta}^{\Delta t'
  \rightarrow t}/N(N-1)$ indicates the mean of the link weights and
$\Delta t'=[1, 5]$. The reciprocity coefficient $\rho_{t}$ ranges from
$-1$ to $1$, and allows us to distinguish between antireciprocal
($\rho_t<0$) and reciprocal ($\rho_t>0$) networks. We have only 
considered in the calculation the top half pairs of mutual links with the largest
weights. 
\bigbreak
\noindent
\emph{(ii) Motifs.} In this study, we focus on three-node motif analysis~\cite{milo2002network}. 
There are 13 different possible connected subgraphs of three nodes.
In order to measure the statistical significance of 
each subgraph $g$, we have computed the $Z$-score $Z_g$ defined as 
\begin{equation}
Z_{g} = (N_{g}-\langle N^{rand}_{g} \rangle)/\sigma_{g}
\label{eq:zscore}
\end{equation}
where $N_{g}$ is the number of times subgraph $g$ appears in the
network, and $\langle N^{rand}_{g} \rangle$ and $\sigma_{g}$ are respectively  
average and standard deviation of the 
number of times subgraph $g$ occurs in an ensemble of randomized
graphs with the same degree distribution as the original network.  
For each year, we have generated an ensemble of $5000$ different randomized
samples of the original network using the configuration model.


\begin{thebibliography}{10}
\urlstyle{rm}
\expandafter\ifx\csname url\endcsname\relax
  \def\url#1{\texttt{#1}}\fi
\expandafter\ifx\csname urlprefix\endcsname\relax\def\urlprefix{URL }\fi
\expandafter\ifx\csname doiprefix\endcsname\relax\def\doiprefix{DOI: }\fi
\providecommand{\bibinfo}[2]{#2}
\providecommand{\eprint}[2][]{\url{#2}}

\bibitem{sekara2018chaperone}
\bibinfo{author}{Sekara, V.} \emph{et~al.}
\newblock \bibinfo{journal}{\bibinfo{title}{The chaperone effect in scientific
  publishing}}.
\newblock {\emph{\JournalTitle{Proceedings of the National Academy of
  Sciences}}} \textbf{\bibinfo{volume}{115}}, \bibinfo{pages}{12603--12607}
  (\bibinfo{year}{2018}).

\bibitem{li2019early}
\bibinfo{author}{Li, W.}, \bibinfo{author}{Aste, T.},
  \bibinfo{author}{Caccioli, F.} \& \bibinfo{author}{Livan, G.}
\newblock \bibinfo{journal}{\bibinfo{title}{Early coauthorship with top
  scientists predicts success in academic careers}}.
\newblock {\emph{\JournalTitle{Nature Communications}}}
  \textbf{\bibinfo{volume}{10}}, \bibinfo{pages}{1--9} (\bibinfo{year}{2019}).

\bibitem{monechi2019efficient}
\bibinfo{author}{Monechi, B.}, \bibinfo{author}{Pullano, G.} \&
  \bibinfo{author}{Loreto, V.}
\newblock \bibinfo{journal}{\bibinfo{title}{Efficient team structures in an
  open-ended cooperative creativity experiment}}.
\newblock {\emph{\JournalTitle{Proceedings of the National Academy of
  Sciences}}} \textbf{\bibinfo{volume}{116}}, \bibinfo{pages}{22088--22093}
  (\bibinfo{year}{2019}).

\bibitem{armano2017beneficial}
\bibinfo{author}{Armano, G.} \& \bibinfo{author}{Javarone, M.~A.}
\newblock \bibinfo{journal}{\bibinfo{title}{The beneficial role of mobility for
  the emergence of innovation}}.
\newblock {\emph{\JournalTitle{Scientific reports}}}
  \textbf{\bibinfo{volume}{7}}, \bibinfo{pages}{1781} (\bibinfo{year}{2017}).

\bibitem{milojevic2018changing}
\bibinfo{author}{Milojevi{\'c}, S.}, \bibinfo{author}{Radicchi, F.} \&
  \bibinfo{author}{Walsh, J.~P.}
\newblock \bibinfo{journal}{\bibinfo{title}{Changing demographics of scientific
  careers: The rise of the temporary workforce}}.
\newblock {\emph{\JournalTitle{Proceedings of the National Academy of
  Sciences}}} \textbf{\bibinfo{volume}{115}}, \bibinfo{pages}{12616--12623}
  (\bibinfo{year}{2018}).

\bibitem{clauset2015systematic}
\bibinfo{author}{Clauset, A.}, \bibinfo{author}{Arbesman, S.} \&
  \bibinfo{author}{Larremore, D.~B.}
\newblock \bibinfo{journal}{\bibinfo{title}{Systematic inequality and hierarchy
  in faculty hiring networks}}.
\newblock {\emph{\JournalTitle{Science advances}}}
  \textbf{\bibinfo{volume}{1}}, \bibinfo{pages}{e1400005}
  (\bibinfo{year}{2015}).

\bibitem{gargiulo2014driving}
\bibinfo{author}{Gargiulo, F.} \& \bibinfo{author}{Carletti, T.}
\newblock \bibinfo{journal}{\bibinfo{title}{Driving forces of researchers
  mobility}}.
\newblock {\emph{\JournalTitle{Scientific reports}}}
  \textbf{\bibinfo{volume}{4}}, \bibinfo{pages}{4860} (\bibinfo{year}{2014}).

\bibitem{deville2014career}
\bibinfo{author}{Deville, P.} \emph{et~al.}
\newblock \bibinfo{journal}{\bibinfo{title}{Career on the move: Geography,
  stratification, and scientific impact}}.
\newblock {\emph{\JournalTitle{Scientific reports}}}
  \textbf{\bibinfo{volume}{4}}, \bibinfo{pages}{4770} (\bibinfo{year}{2014}).

\bibitem{ma2015anatomy}
\bibinfo{author}{Ma, A.}, \bibinfo{author}{Mondrag{\'o}n, R.~J.} \&
  \bibinfo{author}{Latora, V.}
\newblock \bibinfo{journal}{\bibinfo{title}{Anatomy of funded research in
  science}}.
\newblock {\emph{\JournalTitle{Proceedings of the National Academy of
  Sciences}}} \textbf{\bibinfo{volume}{112}}, \bibinfo{pages}{14760--14765}
  (\bibinfo{year}{2015}).

\bibitem{van2015interdisciplinary}
\bibinfo{author}{Van~Noorden, R.}
\newblock \bibinfo{journal}{\bibinfo{title}{Interdisciplinary research by the
  numbers}}.
\newblock {\emph{\JournalTitle{Nature}}} \textbf{\bibinfo{volume}{525}},
  \bibinfo{pages}{306--307} (\bibinfo{year}{2015}).

\bibitem{sinatra2015century}
\bibinfo{author}{Sinatra, R.}, \bibinfo{author}{Deville, P.},
  \bibinfo{author}{Szell, M.}, \bibinfo{author}{Wang, D.} \&
  \bibinfo{author}{Barab{\'a}si, A.-L.}
\newblock \bibinfo{journal}{\bibinfo{title}{A century of physics}}.
\newblock {\emph{\JournalTitle{Nature Physics}}} \textbf{\bibinfo{volume}{11}},
  \bibinfo{pages}{791} (\bibinfo{year}{2015}).

\bibitem{battiston2019taking}
\bibinfo{author}{Battiston, F.} \emph{et~al.}
\newblock \bibinfo{journal}{\bibinfo{title}{Taking census of physics}}.
\newblock {\emph{\JournalTitle{Nature Reviews Physics}}}
  \textbf{\bibinfo{volume}{1}}, \bibinfo{pages}{89--97} (\bibinfo{year}{2019}).

\bibitem{bhagat2002cultural}
\bibinfo{author}{Bhagat, R.~S.}, \bibinfo{author}{Kedia, B.~L.},
  \bibinfo{author}{Harveston, P.~D.} \& \bibinfo{author}{Triandis, H.~C.}
\newblock \bibinfo{journal}{\bibinfo{title}{Cultural variations in the
  cross-border transfer of organizational knowledge: An integrative
  framework}}.
\newblock {\emph{\JournalTitle{Academy of management review}}}
  \textbf{\bibinfo{volume}{27}}, \bibinfo{pages}{204--221}
  (\bibinfo{year}{2002}).

\bibitem{chen2010impact}
\bibinfo{author}{Chen, J.}, \bibinfo{author}{Sun, P.~Y.} \&
  \bibinfo{author}{McQueen, R.~J.}
\newblock \bibinfo{journal}{\bibinfo{title}{The impact of national cultures on
  structured knowledge transfer}}.
\newblock {\emph{\JournalTitle{Journal of knowledge management}}}
  \textbf{\bibinfo{volume}{14}}, \bibinfo{pages}{228--242}
  (\bibinfo{year}{2010}).

\bibitem{bell2007geography}
\bibinfo{author}{Bell, G.~G.} \& \bibinfo{author}{Zaheer, A.}
\newblock \bibinfo{journal}{\bibinfo{title}{Geography, networks, and knowledge
  flow}}.
\newblock {\emph{\JournalTitle{Organization Science}}}
  \textbf{\bibinfo{volume}{18}}, \bibinfo{pages}{955--972}
  (\bibinfo{year}{2007}).

\bibitem{sorenson2006complexity}
\bibinfo{author}{Sorenson, O.}, \bibinfo{author}{Rivkin, J.~W.} \&
  \bibinfo{author}{Fleming, L.}
\newblock \bibinfo{journal}{\bibinfo{title}{Complexity, networks and knowledge
  flow}}.
\newblock {\emph{\JournalTitle{Research policy}}}
  \textbf{\bibinfo{volume}{35}}, \bibinfo{pages}{994--1017}
  (\bibinfo{year}{2006}).

\bibitem{agrawal2008spatial}
\bibinfo{author}{Agrawal, A.}, \bibinfo{author}{Kapur, D.} \&
  \bibinfo{author}{McHale, J.}
\newblock \bibinfo{journal}{\bibinfo{title}{How do spatial and social proximity
  influence knowledge flows? evidence from patent data}}.
\newblock {\emph{\JournalTitle{Journal of urban economics}}}
  \textbf{\bibinfo{volume}{64}}, \bibinfo{pages}{258--269}
  (\bibinfo{year}{2008}).

\bibitem{meyer2002tracing}
\bibinfo{author}{Meyer, M.}
\newblock \bibinfo{journal}{\bibinfo{title}{Tracing knowledge flows in
  innovation systems}}.
\newblock {\emph{\JournalTitle{Scientometrics}}} \textbf{\bibinfo{volume}{54}},
  \bibinfo{pages}{193--212} (\bibinfo{year}{2002}).

\bibitem{acemoglu2016innovation}
\bibinfo{author}{Acemoglu, D.}, \bibinfo{author}{Akcigit, U.} \&
  \bibinfo{author}{Kerr, W.~R.}
\newblock \bibinfo{journal}{\bibinfo{title}{Innovation network}}.
\newblock {\emph{\JournalTitle{Proceedings of the National Academy of
  Sciences}}} \textbf{\bibinfo{volume}{113}}, \bibinfo{pages}{11483--11488}
  (\bibinfo{year}{2016}).

\bibitem{zeng2017science}
\bibinfo{author}{Zeng, A.} \emph{et~al.}
\newblock \bibinfo{journal}{\bibinfo{title}{The science of science: From the
  perspective of complex systems}}.
\newblock {\emph{\JournalTitle{Physics Reports}}}
  \textbf{\bibinfo{volume}{714}}, \bibinfo{pages}{1--73}
  (\bibinfo{year}{2017}).

\bibitem{zhang2013characterizing}
\bibinfo{author}{Zhang, Q.}, \bibinfo{author}{Perra, N.},
  \bibinfo{author}{Gon{\c{c}}alves, B.}, \bibinfo{author}{Ciulla, F.} \&
  \bibinfo{author}{Vespignani, A.}
\newblock \bibinfo{journal}{\bibinfo{title}{Characterizing scientific
  production and consumption in physics}}.
\newblock {\emph{\JournalTitle{Scientific reports}}}
  \textbf{\bibinfo{volume}{3}}, \bibinfo{pages}{1640} (\bibinfo{year}{2013}).

\bibitem{borner2006mapping}
\bibinfo{author}{B{\"o}rner, K.}, \bibinfo{author}{Penumarthy, S.},
  \bibinfo{author}{Meiss, M.} \& \bibinfo{author}{Ke, W.}
\newblock \bibinfo{journal}{\bibinfo{title}{Mapping the diffusion of scholarly
  knowledge among major us research institutions}}.
\newblock {\emph{\JournalTitle{Scientometrics}}} \textbf{\bibinfo{volume}{68}},
  \bibinfo{pages}{415--426} (\bibinfo{year}{2006}).

\bibitem{zhuge2002knowledge}
\bibinfo{author}{Zhuge, H.}
\newblock \bibinfo{journal}{\bibinfo{title}{A knowledge flow model for
  peer-to-peer team knowledge sharing and management}}.
\newblock {\emph{\JournalTitle{Expert systems with applications}}}
  \textbf{\bibinfo{volume}{23}}, \bibinfo{pages}{23--30}
  (\bibinfo{year}{2002}).

\bibitem{yan2016disciplinary}
\bibinfo{author}{Yan, E.}
\newblock \bibinfo{journal}{\bibinfo{title}{Disciplinary knowledge production
  and diffusion in science}}.
\newblock {\emph{\JournalTitle{Journal of the Association for Information
  Science and Technology}}} \textbf{\bibinfo{volume}{67}},
  \bibinfo{pages}{2223--2245} (\bibinfo{year}{2016}).

\bibitem{latora2017complex}
\bibinfo{author}{Latora, V.}, \bibinfo{author}{Nicosia, V.} \&
  \bibinfo{author}{Russo, G.}
\newblock \emph{\bibinfo{title}{Complex networks: principles, methods and
  applications}} (\bibinfo{publisher}{Cambridge University Press},
  \bibinfo{year}{2017}).

\bibitem{newman2018networks}
\bibinfo{author}{Newman, M.}
\newblock \emph{\bibinfo{title}{Networks}} (\bibinfo{publisher}{Oxford
  university press}, \bibinfo{year}{2018}).

\bibitem{pan2012evolution}
\bibinfo{author}{Pan, R.~K.}, \bibinfo{author}{Sinha, S.},
  \bibinfo{author}{Kaski, K.} \& \bibinfo{author}{Saram{\"a}ki, J.}
\newblock \bibinfo{journal}{\bibinfo{title}{The evolution of
  interdisciplinarity in physics research}}.
\newblock {\emph{\JournalTitle{Scientific reports}}}
  \textbf{\bibinfo{volume}{2}}, \bibinfo{pages}{551} (\bibinfo{year}{2012}).

\bibitem{bonaventura2017advantages}
\bibinfo{author}{Bonaventura, M.}, \bibinfo{author}{Latora, V.},
  \bibinfo{author}{Nicosia, V.} \& \bibinfo{author}{Panzarasa, P.}
\newblock \bibinfo{journal}{\bibinfo{title}{The advantages of
  interdisciplinarity in modern science}}.
\newblock {\emph{\JournalTitle{arXiv preprint arXiv:1712.07910}}}
  (\bibinfo{year}{2017}).

\bibitem{pluchino2019exploring}
\bibinfo{author}{Pluchino, A.} \emph{et~al.}
\newblock \bibinfo{journal}{\bibinfo{title}{Exploring the role of
  interdisciplinarity in physics: Success, talent and luck}}.
\newblock {\emph{\JournalTitle{PloS one}}} \textbf{\bibinfo{volume}{14}},
  \bibinfo{pages}{e0218793} (\bibinfo{year}{2019}).

\bibitem{tria2014dynamics}
\bibinfo{author}{Tria, F.}, \bibinfo{author}{Loreto, V.},
  \bibinfo{author}{Servedio, V. D.~P.} \& \bibinfo{author}{Strogatz, S.~H.}
\newblock \bibinfo{journal}{\bibinfo{title}{The dynamics of correlated
  novelties}}.
\newblock {\emph{\JournalTitle{Scientific reports}}}
  \textbf{\bibinfo{volume}{4}}, \bibinfo{pages}{5890} (\bibinfo{year}{2014}).

\bibitem{iacopini2018innovation}
\bibinfo{author}{Iacopini, I.},
  \bibinfo{author}{Milojevi\ifmmode~\acute{c}\else \'{c}\fi{}, S. c.~v.} \&
  \bibinfo{author}{Latora, V.}
\newblock \bibinfo{journal}{\bibinfo{title}{Network dynamics of innovation
  processes}}.
\newblock {\emph{\JournalTitle{Phys. Rev. Lett.}}}
  \textbf{\bibinfo{volume}{120}}, \bibinfo{pages}{048301}
  (\bibinfo{year}{2018}).

\bibitem{chinazzi2019mapping}
\bibinfo{author}{Chinazzi, M.}, \bibinfo{author}{Gon{\c{c}}alves, B.},
  \bibinfo{author}{Zhang, Q.} \& \bibinfo{author}{Vespignani, A.}
\newblock \bibinfo{journal}{\bibinfo{title}{Mapping the physics research space:
  a machine learning approach}}.
\newblock {\emph{\JournalTitle{EPJ Data Science}}}
  \textbf{\bibinfo{volume}{8}}, \bibinfo{pages}{33} (\bibinfo{year}{2019}).

\bibitem{zhu2003effect}
\bibinfo{author}{Zhu, H.}, \bibinfo{author}{Wang, X.} \& \bibinfo{author}{Zhu,
  J.-Y.}
\newblock \bibinfo{journal}{\bibinfo{title}{Effect of aging on network
  structure}}.
\newblock {\emph{\JournalTitle{Physical Review E}}}
  \textbf{\bibinfo{volume}{68}}, \bibinfo{pages}{056121}
  (\bibinfo{year}{2003}).

\bibitem{milo2002network}
\bibinfo{author}{Milo, R.} \emph{et~al.}
\newblock \bibinfo{journal}{\bibinfo{title}{Network motifs: simple building
  blocks of complex networks}}.
\newblock {\emph{\JournalTitle{Science}}} \textbf{\bibinfo{volume}{298}},
  \bibinfo{pages}{824--827} (\bibinfo{year}{2002}).

\bibitem{tahamtan2016factors}
\bibinfo{author}{Tahamtan, I.}, \bibinfo{author}{Afshar, A.~S.} \&
  \bibinfo{author}{Ahamdzadeh, K.}
\newblock \bibinfo{journal}{\bibinfo{title}{Factors affecting number of
  citations: a comprehensive review of the literature}}.
\newblock {\emph{\JournalTitle{Scientometrics}}}
  \textbf{\bibinfo{volume}{107}}, \bibinfo{pages}{1195--1225}
  (\bibinfo{year}{2016}).

\bibitem{cirkel2008history}
\bibinfo{author}{Cirkel-Bartelt, V.}
\newblock \bibinfo{journal}{\bibinfo{title}{History of astroparticle physics
  and its components}}.
\newblock {\emph{\JournalTitle{Living Reviews in Relativity}}}
  \textbf{\bibinfo{volume}{11}}, \bibinfo{pages}{2} (\bibinfo{year}{2008}).

\bibitem{rinia2002measuring}
\bibinfo{author}{Rinia, E.~J.}, \bibinfo{author}{Van~Leeuwen, T.~N.},
  \bibinfo{author}{Bruins, E.~E.}, \bibinfo{author}{Van~Vuren, H.~G.} \&
  \bibinfo{author}{Van~Raan, A.~F.}
\newblock \bibinfo{journal}{\bibinfo{title}{Measuring knowledge transfer
  between fields of science}}.
\newblock {\emph{\JournalTitle{Scientometrics}}} \textbf{\bibinfo{volume}{54}},
  \bibinfo{pages}{347--362} (\bibinfo{year}{2002}).

\bibitem{phene2006breakthrough}
\bibinfo{author}{Phene, A.}, \bibinfo{author}{Fladmoe-Lindquist, K.} \&
  \bibinfo{author}{Marsh, L.}
\newblock \bibinfo{journal}{\bibinfo{title}{Breakthrough innovations in the us
  biotechnology industry: the effects of technological space and geographic
  origin}}.
\newblock {\emph{\JournalTitle{Strategic Management Journal}}}
  \textbf{\bibinfo{volume}{27}}, \bibinfo{pages}{369--388}
  (\bibinfo{year}{2006}).

\bibitem{garlaschelli2004patterns}
\bibinfo{author}{Garlaschelli, D.} \& \bibinfo{author}{Loffredo, M.~I.}
\newblock \bibinfo{journal}{\bibinfo{title}{Patterns of link reciprocity in
  directed networks}}.
\newblock {\emph{\JournalTitle{Physical review letters}}}
  \textbf{\bibinfo{volume}{93}}, \bibinfo{pages}{268701}
  (\bibinfo{year}{2004}).
\end{thebibliography}

\section*{Acknowledgements}
This work was funded by the Leverhulme Trust Research Fellowship "CREATE: the network components of creativity and success".

\section*{Author contributions statement}
Y.S. and V.L. designed research; Y.S. performed research; Y.S. and V.L. analyzed data; and Y.S and V.L. wrote the paper.

\section*{Additional information}
\textbf{Competing interests} The authors declare no competing financial interests.



\end{document}